\documentclass[conference]{IEEEtran}
\IEEEoverridecommandlockouts
\usepackage{cite}
\usepackage{amsmath,amssymb,amsfonts}
\usepackage{algorithmic}
\usepackage{graphicx}
\usepackage{textcomp}
\usepackage{xcolor}
\usepackage[T1]{fontenc}
\usepackage[utf8]{inputenc}
\usepackage[english]{babel}
\usepackage[breaklinks]{hyperref}
\usepackage{booktabs}
\usepackage{array}
\usepackage{multirow}
\usepackage{geometry}
\usepackage{microtype}

\def\BibTeX{{\rm B\kern-.05em{\sc i\kern-.025em b}\kern-.08em
    T\kern-.1667em\lower.7ex\hbox{E}\kern-.125emX}}

\begin{document}

\title{Implementing Keyword Spotting on the MCXN947 Microcontroller with Integrated NPU\\
{}

}

\author{
\IEEEauthorblockN{1\textsuperscript{st} Petar Jakuš}
\IEEEauthorblockA{
\textit{Faculty of Electrical Engineering and Computing}\\
Zagreb, Croatia \\
petar.jakus@fer.unizg.hr}

\and  

\IEEEauthorblockN{2\textsuperscript{nd} Hrvoje Džapo}
\IEEEauthorblockA{
\textit{Faculty of Electrical Engineering and Computing}\\
Zagreb, Croatia \\
hrvoje.dzapo@fer.unizg.hr}
}

\maketitle

\begin{abstract}
This paper presents a keyword spotting (KWS) system implemented on the NXP MCXN947 microcontroller with an integrated Neural Processing Unit (NPU), enabling real-time voice interaction on resource-constrained devices. The system combines MFCC feature extraction with a CNN classifier, optimized using Quantization Aware Training to reduce model size with minimal accuracy drop. Experimental results demonstrate a 59$\times$ speedup in inference time when leveraging the NPU compared to CPU-only execution, achieving 97.06\% accuracy with a model size of 30.58 KB, demonstrating the feasibility of efficient, low-power voice interfaces on embedded platforms.

\end{abstract}


\section{Introduction}

With the growing adoption of IoT devices, keyword spotting (KWS) has become essential for voice-controlled interaction in smart assistants and embedded systems. However, implementing KWS on resource-constrained devices remains challenging due to limited processing power, memory constraints, and strict energy requirements, demanding optimized solutions for microcontroller deployment. A typical keyword spotting system consists of signal acquisition, feature extraction and a neural network \cite{gruenstein2017cascade}.

Signal acquisition in keyword spotting systems is typically performed using a microphone, most often a MEMS microphone, which converts pressure variations into electrical signals. The raw audio signal often contains noise and amplitude variations, requiring signal preprocessing to improve robustness and accuracy. Common noise reduction techniques include adaptive noise cancellation (ANC) \cite{8462346} and beamforming \cite{beamformares}. Feature extraction transforms the input signal into a compact and informative representation suitable for classification. Widely used methods include Mel-frequency cepstral coefficients (MFCC) \cite{ali2020mel,kundu2023heimdal} and RASTA-PLP \cite{zeng2006rasta,hermansky1991rasta}. To reduce the computational complexity of the preprocessing stage, simplified approaches such as direct application of the Fast Fourier Transform (FFT) have been proposed \cite{amoh2019optimized}. For ultra-low-power applications, analog front-end approaches and featureless techniques have been explored \cite{cerutti2022sub,yang2018afenn,ulkar2021ultra}. Alternatively, feature extraction can be integrated into the neural network itself, as shown in \cite{vitolo2023automatic,chen2015query}.

The classification stage typically employs neural network architectures. Convolutional Neural Networks (CNNs) have shown strong performance in KWS tasks, especially when combined with model compression techniques such as pruning and quantization \cite{leroy2019federated,cioflan2024device,zhang2017hello}. Hardware acceleration for neural network inference, such as using dedicated co-processors, has been shown to significantly reduce power consumption \cite{ulkar2021ultra}. Recurrent architectures like Long Short-Term Memory (LSTM) networks \cite{giraldo2018lstmaccelerator} and Gated Recurrent Units (GRUs), including energy-efficient variants such as eGRU \cite{amoh2019optimized}, are also employed to capture temporal dependencies in speech signals. Advanced compression techniques such as network binarization have further reduced memory footprints for deployment on edge devices \cite{cerutti2022sub}. Hybrid architectures combining CNNs and RNNs have been proposed to improve classification performance while maintaining efficiency \cite{zhang2017hello}.

This work presents an optimized KWS system targeting the NXP MCXN947 microcontroller, leveraging its integrated NPU to accelerate a quantized CNN model with MFCC-based features. By employing quantization-aware training (QAT), the method minimizes accuracy loss while ensuring efficient quantization, significantly reducing model size and inference latency. Experimental results demonstrate the feasibility of deploying responsive, low-power voice interfaces on resource-constrained edge devices.

\section{Methodology}

\subsection{Data}
The Google Speech Commands dataset\cite{warden2018speechcommandsdatasetlimitedvocabulary} contains 105,829 one-second voice recordings of 35 words, recorded at 16 kHz sample rate on mobile devices. The dataset is divided into training (84,843), test (11,005), and validation (9,981) sets as originally proposed.

\subsubsection{\textbf{Preprocessing}}
\label{sec:preprocessing}
Mel-Frequency Cepstral Coefficients (MFCC) were used for feature extraction through the following pipeline:
\begin{itemize}
    \item \textbf{Framing and Windowing:} Audio signals were segmented into 25 ms frames with 10 ms hop size (400 samples at 16 kHz), as illustrated in Figure~\ref{fig:mfcc_framing}. A Hamming window was applied to each frame to reduce spectral leakage (Figure~\ref{fig:hamming_frame}):
    \[
    w(n) = 0.54 - 0.46 \cdot \cos\left(\frac{2\pi n}{N}\right)
    \]
    
\begin{figure}
\centering
\includegraphics[width=1\linewidth]{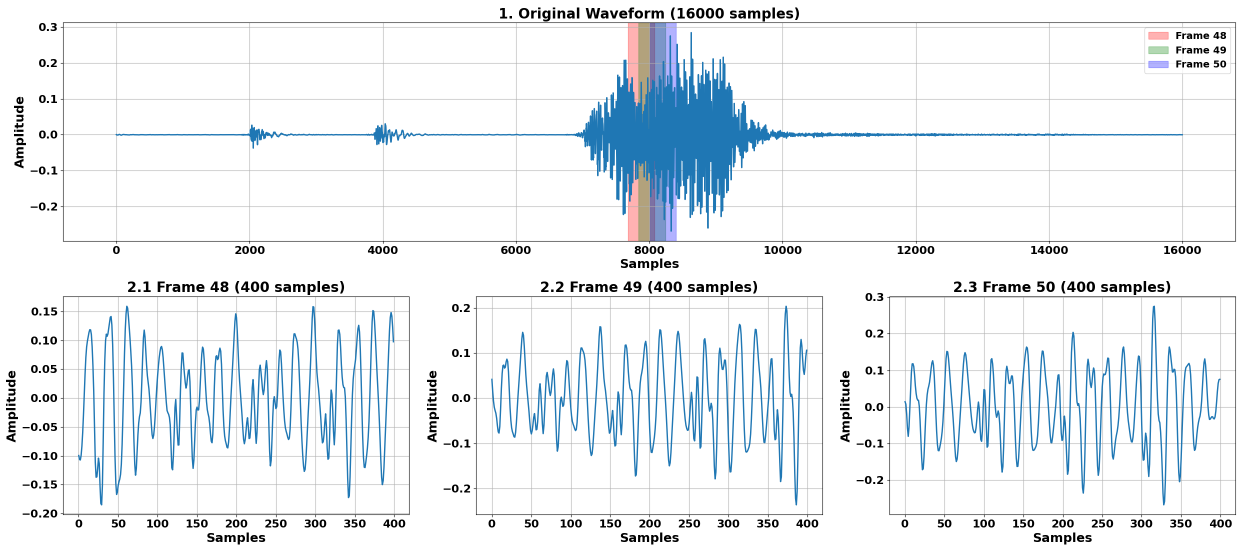}
\caption{MFCC original signal and frames.}
\label{fig:mfcc_framing}
\end{figure}

\begin{figure}
\centering
\includegraphics[width=1\linewidth]{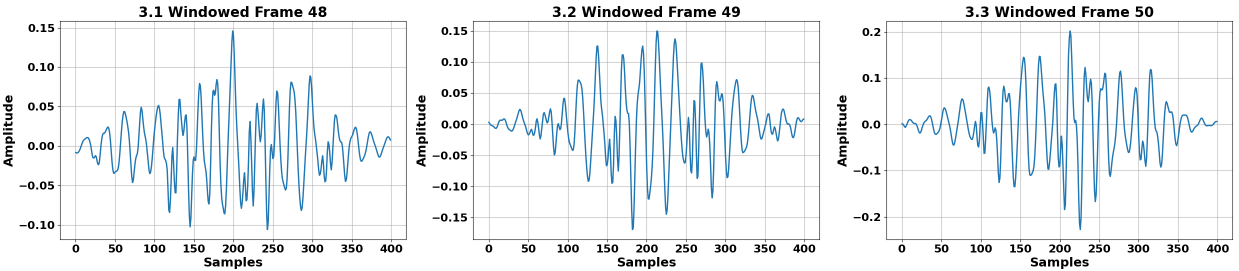}
\caption{Frames after applying Hamming function.}
\label{fig:hamming_frame}
\end{figure}
    
    \item \textbf{Spectral Analysis:} Fast Fourier Transform (FFT) converted each windowed frame to the frequency domain, followed by power spectrum calculation (Figure~\ref{fig:power}). For a given FFT output $X[k]$, the power spectrum $P[k]$ is defined as:
    \[
    P[k] = \frac{1}{N} |X[k]|^2
    \]
    where $N$ is the FFT length.

\begin{figure}
\centering
\includegraphics[width=1\linewidth]{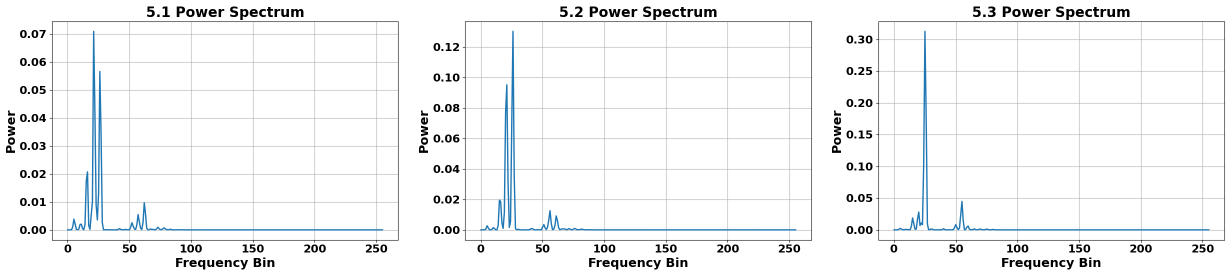}
\caption{Power spectrum.}
\label{fig:power}
\end{figure}

    \item \textbf{Mel-scale Filtering:} A bank of 40 filters spanning 40 Hz to 7.6 kHz was applied to align with human auditory perception and \ref{fig:mel_spectrum}). The Mel scale provides higher frequency resolution at lower frequencies, matching human hearing characteristics.

\begin{figure}
\centering
\includegraphics[width=1\linewidth]{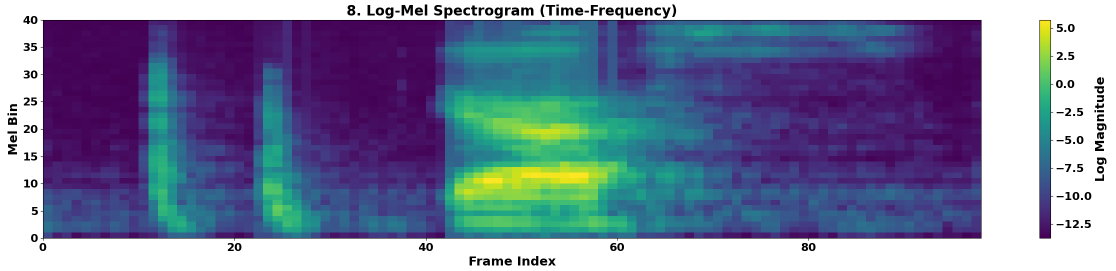}
\caption{Frames Mel spectrums.}
\label{fig:mel_spectrum}
\end{figure}

    \item \textbf{MFCC Computation:} Discrete Cosine Transform (DCT) was applied to the log mel-spectrum to decorrelate coefficients, producing 20 MFCC features per frame following HTK convention (Figure~\ref{fig:mfcc}).

\begin{figure}
\centering
\includegraphics[width=1\linewidth]{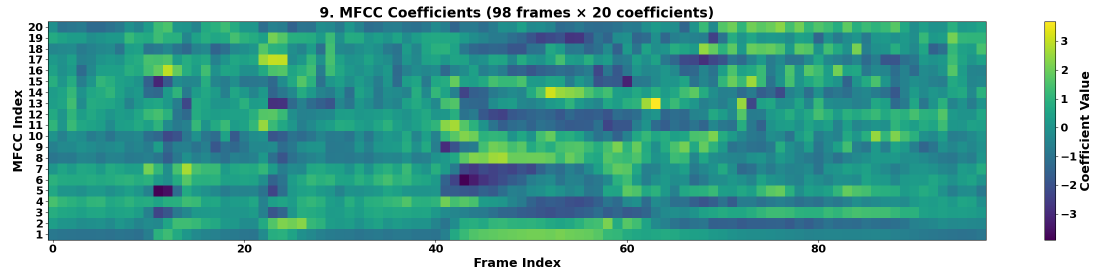}
\caption{MFCC spectogram.}
\label{fig:mfcc}
\end{figure}

\end{itemize}

This preprocessing pipeline transforms raw audio into perceptually-motivated features suitable for speech recognition tasks.

\subsection{Model Architecture and Quantization}

The proposed architecture employs a compact CNN optimized for edge deployment, featuring two convolutional layers (each with 2D convolution, batch normalization, and max pooling) followed by two dense layers. The model is trained specifically for the "Marvin" keyword using Adam optimization (learning rate = 0.001) over 10 epochs, with class weights of 24.81 (Marvin) and 0.51 (non-Marvin) to address imbalance.

The model architecture incorporates quantization-aware training (QAT) to enable efficient conversion from 32-bit floating-point to 8-bit fixed-point weight representation. The complete QAT approach applies quantization constraints to all layers during training, simulating 8-bit operations in forward passes while maintaining full precision for backward propagation. While this increases training time by approximately 20\%, it ensures minimal accuracy loss when after quantization. Alternatively, a fine-tuning approach can be applied where a pre-trained full-precision model is adapted with QAT for fewer epochs, offering reduced training time with comparable accuracy.

The regular model architecture, the QAT prepared version and final quantized implementation are detailed in Tables~\ref{tab:original_model}, \ref{tab:qat_model}, and \ref{tab:quantized_model}, respectively, including layer-specific quantization parameters.

\begin{table}[h!]
\centering
\caption{Regular Neural Network Architecture}
\label{tab:original_model}
\begin{tabular}{@{}lcc@{}}
\toprule
\textbf{Layer (type)} & \textbf{Output Shape} & \textbf{Param \#} \\
\midrule
input (InputLayer) & (None, 98, 20, 1) & 0 \\
conv1 (Conv2D) & (None, 96, 18, 32) & 320 \\
bn1 (BatchNormalization) & (None, 96, 18, 32) & 128 \\
pool1 (MaxPooling2D) & (None, 48, 9, 32) & 0 \\
conv2 (Conv2D) & (None, 46, 7, 64) & 18496 \\
bn2 (BatchNormalization) & (None, 46, 7, 64) & 256 \\
pool2 (MaxPooling2D) & (None, 23, 3, 64) & 0 \\
gap (GlobalAveragePooling2D) & (None, 64) & 0 \\
dropout (Dropout) & (None, 64) & 0 \\
fc1 (Dense) & (None, 128) & 8320 \\
output (Dense) & (None, 1) & 129 \\
\midrule
\textbf{Total params:} & \multicolumn{2}{c}{\textbf{27649 (108.00 KB)}} \\
\textbf{Trainable params:} & \multicolumn{2}{c}{\textbf{27457 (107.25 KB)}} \\
\textbf{Non-trainable params:} & \multicolumn{2}{c}{\textbf{192 (768.00 B)}} \\
\bottomrule
\end{tabular}
\end{table}

\begin{table}[h!]
\centering
\caption{QAT Prepared Neural Network Architecture}
\label{tab:qat_model}
\begin{tabular}{@{}lcc@{}}
\toprule
\textbf{Layer (type)} & \textbf{Output Shape} & \textbf{Param \#} \\
\midrule
input (InputLayer) & (None, 98, 20, 1) & 0 \\
quantize\_layer (QuantizeLayer) & (None, 98, 20, 1) & 3 \\
quant\_conv1 (QuantizeWrapperV2) & (None, 96, 18, 32) & 387 \\
quant\_bn1 (QuantizeWrapperV2) & (None, 96, 18, 32) & 129 \\
quant\_pool1 (QuantizeWrapperV2) & (None, 48, 9, 32) & 1 \\
quant\_conv2 (QuantizeWrapperV2) & (None, 46, 7, 64) & 18627 \\
quant\_bn2 (QuantizeWrapperV2) & (None, 46, 7, 64) & 257 \\
pool2 (MaxPooling2D) & (None, 23, 3, 64) & 0 \\
gap (GlobalAveragePooling2D) & (None, 64) & 0 \\
quant\_dropout (QuantizeWrapperV2) & (None, 64) & 1 \\
quant\_fc1 (QuantizeWrapperV2) & (None, 128) & 8325 \\
quant\_output (QuantizeWrapperV2) & (None, 1) & 134 \\
\midrule
\textbf{Total params:} & \multicolumn{2}{c}{\textbf{27864 (108.84 KB)}} \\
\textbf{Trainable params:} & \multicolumn{2}{c}{\textbf{27457 (107.25 KB)}} \\
\textbf{Non-trainable params:} & \multicolumn{2}{c}{\textbf{407 (1.59 KB)}} \\
\bottomrule
\end{tabular}
\end{table}

\begin{table}[h!]
\centering
\caption{Quantized Model Architecture}
\label{tab:quantized_model}
\begin{tabular}{@{}lll@{}}
\toprule
\textbf{Layer (type)} & \textbf{Output Shape} & \textbf{Param \#} \\
\midrule
\multirow{2}{*}{InputLayer} & [-1, 98, 20, 1] & 0\\
\addlinespace
\multirow{2}{*}{Conv2D} & \multirow{2}{*}{[-1, 96, 18, 32]} & 320 weights \\ 
 & & + 32 biases \\
 \addlinespace
\multirow{2}{*}{MUL, ADD (BatchNorm)} & \multirow{2}{*}{[-1, 96, 18, 32]} & 64 scale \\ 
 & & + 32 offset \\
 \addlinespace
\multirow{2}{*}{MaxPool2D} & [-1, 48, 9, 32] & 0 \\
\addlinespace
\multirow{2}{*}{Conv2D} & \multirow{2}{*}{[-1, 46, 7, 64]} & 18,432 weights \\ 
 & & + 64 biases \\
 \addlinespace
\multirow{2}{*}{MUL, ADD (BatchNorm)} & \multirow{2}{*}{[-1, 46, 7, 64]} & 64 scale \\ 
 & & + 64 offset \\
 \addlinespace
MaxPool2D & [-1, 23, 3, 64] & 0 \\
\addlinespace
Mean & [-1, 64] & 0 \\
\addlinespace
\multirow{2}{*}{FullyConnected} & \multirow{2}{*}{[-1, 128]} & 8,192 weights \\ 
 & & + 128 biases \\ 
\addlinespace
\multirow{2}{*}{FullyConnected} & \multirow{2}{*}{[-1, 1]} & 128 weights \\ 
 & & + 1 bias \\ 
\addlinespace
Logistic & [-1, 1] & 0 \\

\midrule
\textbf{Total params:} & \multicolumn{2}{c}{\textbf{27521 (28.20 KB)}} \\
\textbf{Trainable params:} & \multicolumn{2}{c}{\textbf{27297 (27.97 KB)}} \\
\textbf{Non-trainable params:} & \multicolumn{2}{c}{\textbf{224 (224 B)}} \\
\bottomrule
\end{tabular}
\end{table}

\subsection{Model deployment}
\begin{figure}[h]
\centering
\includegraphics[width=0.3\linewidth]{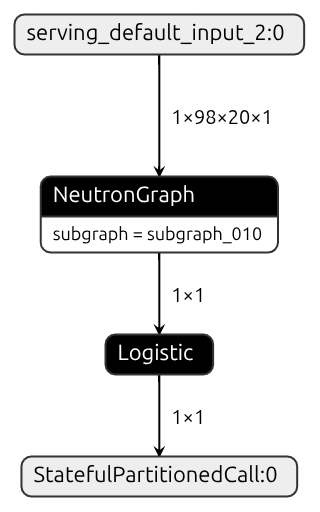}
\caption{Model architecture after NPU conversion \cite{netron}.}
\label{fig:conv_model}
\end{figure}

The quantized TensorFlow Lite model was deployed on the MCXN947 microcontroller both using its eIQ Neutron NPU accelerator and the ARM Cortex-M33 CPU. The quantized model was converted to an NPU-compatible format through NXP's eIQ Toolkit. As shown in Figure~\ref{fig:conv_model}, the conversion process restructured certain layers to take advantage of the dedicated hardware acceleration capabilities of the NPU. Both models are converted to static arrays and stored in flash memory of the MCU.

\section{Results} 

The MFCC feature extraction demonstrated efficient processing with a average computation time of 431 \textmu s per frame on Cortex-M33. Evaluation of three variants of the model revealed significant performance differences.

The regular floating-point model achieved 99.14\% accuracy, with the confusion matrix shown in Table~\ref{tab:regular_cm}. On the Intel Core i5-10210U CPU, this model required 58.67 ms per inference with a memory footprint of 383,674 bytes.

Quantization reduced the size of the model to 35,744 bytes (reduction of 90.68\%) while achieving accuracy of 97.06\%, with the confusion matrix shown in Table~\ref{tab:qantized_cm}. The quantized version showed faster inference at 0.42 ms on the same Intel CPU. When deployed on the MCXN947's Cortex-M33, the inference time increased to 228.2 ms.

\begin{table}[h!]
\centering
\caption{Regular model confusion matrix}
\label{tab:regular_cm}
\begin{tabular}{@{}lcc@{}}
\toprule
\textbf{} & \textbf{Predicted Non-Marvin} & \textbf{Predicted Marvin} \\
\midrule
\textbf{Real Non-Marvin} & 10746 & 64 \\
\textbf{Real Marvin} & 31 & 164 \\
\bottomrule
\end{tabular}
\end{table}

\begin{table}[h!]
\centering
\caption{Quantized model confusion matrix}
\label{tab:qantized_cm}
\begin{tabular}{@{}lcc@{}}
\toprule
\textbf{} & \textbf{Predicted Non-Marvin} & \textbf{Predicted Marvin} \\
\midrule
\textbf{Real Non-Marvin} & 10501 & 309 \\
\textbf{Real Marvin} & 15 & 180 \\
\bottomrule
\end{tabular}
\end{table}

The NPU-optimized version preserved the quantized model's 97.06\% accuracy while further reducing both model size (30,576 bytes) and inference time (3,847 \textmu s on MCXN947). This represents a 98.3\% reduction in model size and 59$\times$ speedup compared to the implementation using only the ARM Cortex-M33 core. The complete performance metrics are summarized in Table~\ref{tab:statistic}.

\begin{table}[h!]
\centering
\caption{Model statistics}
\label{tab:statistic}
\begin{tabular}{@{}lcccc@{}}
\toprule
                   & \multicolumn{1}{l}{\textbf{Size {[}B{]}}} & \multicolumn{1}{l}{\textbf{Acc. {[}\%{]}}} & \multicolumn{1}{l}{\textbf{i5 {[}ms{]}}} & \multicolumn{1}{l}{\textbf{MCXN947 {[}\textmu s{]}}} \\ \midrule
\textbf{Regular}   & 383674                                    & 99.14                                      & 50.67                                    & -                                             \\
\textbf{Quantized} & 35744                                     & 97.06                                      & 0.42                                     & 228210                                        \\
\textbf{NPU}       & 30576                                     & 97.06                                      & -                                        & 3847                                          \\ \bottomrule
\end{tabular}
\end{table}

\section{Conclusion}

This paper demonstrates the successful implementation of keyword spotting on the NXP MCXN947 microcontroller using quantization-aware training and NPU acceleration. The approach achieves a 90.68\% reduction in model size with 97.06\% accuracy, while NPU acceleration provides a 59$\times$ speedup compared to CPU execution. The complete processing pipeline requires less than 5 ms, enabling deployment on embedded platforms.

The results validate the feasibility of implementing a speech recognition system in resource-constrained devices through the combination of model compression and specialized hardware acceleration. This work contributes quantifiable performance metrics for edge AI deployment in privacy-preserving and energy-constrained applications.

Future work should investigate power consumption characteristics, multi-keyword architectures, and robustness evaluation under varying acoustic conditions.
\bibliography{lit}
\bibliographystyle{plain}

\end{document}